\pdfoutput=1
\documentclass[11pt]{article}
\usepackage{microtype}
\usepackage[final]{acl}

\usepackage{multirow}
\usepackage{times}
\usepackage{latexsym}
\usepackage{booktabs} 
\usepackage{amsmath}
\usepackage{makecell}
\usepackage{float}
\usepackage{adjustbox}
\usepackage{xcolor}
\usepackage[T1]{fontenc}
\usepackage[utf8]{inputenc}

\usepackage{microtype}
\usepackage{algorithm}
\usepackage{algorithmic}
\usepackage{amsmath}
\usepackage{amsfonts}
\usepackage{inconsolata}

\usepackage{graphicx}
\usepackage{colortbl}
\title{InfoGain-RAG: Boosting Retrieval-Augmented Generation via Document Information Gain-based Reranking and Filtering}

\author{
    \textbf{Zihan Wang\textsuperscript{1,\thanks{Equal Contribution.}}}\ \ \ \ \ 
    \textbf{Zihan Liang\textsuperscript{1,\footnotemark[1]}}\ \ \ \ \ 
    \textbf{Zhou Shao\textsuperscript{2,\footnotemark[1]}}\ \ \ \ \ 
    \textbf{Yufei Ma\textsuperscript{1}}
    \\
    \textbf{Huangyu Dai\textsuperscript{1}}\ \ \ \ \ 
    \textbf{Ben Chen\textsuperscript{1,\thanks{Corresponding Author.}}}\ \ \ \ \ 
    \textbf{Lingtao Mao\textsuperscript{1}}\ \ \ \ \ 
    \textbf{Chenyi Lei\textsuperscript{1}} 
    \\
    \textbf{Yuqing Ding\textsuperscript{1}}\ \ \ \ \ 
    \textbf{Han Li\textsuperscript{1}}\ \ \ \ \ 
    \\
    \textsuperscript{1}{Kuaishou Technology, Beijing, China} \\
    \textsuperscript{2}{Peking University, Beijing, China} \\
    \\
}

\begin{document}
\maketitle
\begin{abstract}

Retrieval-Augmented Generation (RAG) has emerged as a promising approach to address key limitations of Large Language Models (LLMs), such as hallucination, outdated knowledge, and lacking reference. However, current RAG frameworks often struggle with identifying whether retrieved documents meaningfully contribute to answer generation. This shortcoming makes it difficult to filter out irrelevant or even misleading content, which notably impacts the final performance.
In this paper, we propose Document Information Gain (DIG), a novel metric designed to quantify the contribution of retrieved documents to correct answer generation. DIG measures a document's value by computing the difference of LLM's generation confidence with and without the document augmented. 
Further, we introduce InfoGain-RAG, a framework that leverages DIG scores to train a specialized reranker, which prioritizes each retrieved document from exact distinguishing and accurate sorting perspectives. This approach can effectively filter out irrelevant documents and select the most valuable ones for better answer generation.
Extensive experiments across various models and benchmarks demonstrate that InfoGain-RAG can significantly outperform existing approaches, on both single and multiple retrievers paradigm. Specifically on NaturalQA, it achieves the improvements of 17.9\%, 4.5\%, 12.5\% in exact match accuracy against naive RAG, self-reflective RAG and modern ranking-based RAG respectively, and even an average of 15.3\% increment on advanced proprietary model GPT-4o across all datasets.
These results demonstrate the feasibility of InfoGain-RAG as it can offer a reliable solution for RAG in multiple applications.
\end{abstract}

\section{Introduction}
\par
Recent advancements in Natural Language Processing (NLP) have been significantly propelled by the emergence of LLMs \cite{brown2020language,openai2024gpt4}, which demonstrates remarkable capabilities across many knowledge-intensive tasks. 
However, maintaining reliability remains an ongoing challenge for LLMs, as they often struggle with issues such as hallucination, outdated information and lacking reference.
RAG has emerged as a promising solution to the aforementioned issues. It can enhance responses by augmenting prompts with external information, especially when the model's inherent knowledge is limited \cite{ram2023incontext}. 
However, the generation quality heavily depends on both the selection of relevant documents and their sequential ordering within the LLMs' context window \cite{liu2023lost}.

\par
Research addressing RAG document prioritization spans multiple perspectives, of which three pipelines gain significant attention. \cite{test1,test2,test3} The first pipeline focuses on retriever optimization, which enhances retrieval performance through task-specific training \cite{lewis2020rag,shi2023replug,pkdd}. However, this approach becomes impractical when working with multiple retrievers \cite{rag-survey}. 
The second pipeline leverages LLMs' self-reflection capabilities to evaluate the utility of documents. It employs LLMs to analyze each document and determine whether it should be used. Although feasible, the multiple LLM calls introduce substantial computational overhead ~\cite{selfrag,corrective-rag,main-rag}. 
The third pipeline adds a reranker after the retrieval stage to reorder all retrieved documents \cite{bge-reranker,rag_with_bge}. While this approach can effectively address multiple retrievers, the only consideration on semantic similarity may fail to select the most useful documents for generation (as shown in the Figure \ref{case-reranker} of Appendix \ref{Appendix A}). All these shortcomings limit their further practical application.

\par
To address these limitations, we propose a novel RAG framework, InfoGain-RAG, to filter out irrelevant or even misleading documents, and prioritize the most valuable ones for answer generation. 
Specifically, we firstly introduce a new metric named Document Information Gain (DIG), which calculates the change in LLM's generation confidence with and without the document augmented. A higher DIG score means the document has higher information value.  Then, a multi-task training strategy is designed, enabling one newly added reranking module to predict the DIG score for each document. Only those with a score greater than a certain threshold will be augmented into the LLM for final generation. 
This reranking module is plug-and-play across diverse models and tasks. 
Furthermore, it can efficiently handle documents from multiple retrievers by invoking LLM only once for the entire process and the low computational overhead makes it feasible for the real application.

\par
Extensive evaluations on two different types of tasks: open-domain question answering (TriviaQA \cite{joshi2017triviaqa}, NaturalQA \cite{kwiatkowski2019natural}, and PopQA \cite{popqa}) and fact verification (FM2 \cite{fm2}) spanning both proprietary LLMs (GPT, Claude) \cite{chatgpt,claude3} and open-source models (LLaMA, Qwen, Gemma, DeepSeek) \cite{llama2,qwen,gemma,deepseek}, demonstrate substantial improvements of InfoGain-RAG over existing methods. 
Specifically on NaturalQA, it achieves significant gains in Exact Match accuracy: outperforming naive RAG by 17.9\%, retriever-optimized RAG by 6.8\%, self-reflective RAG by 4.5\%, and modern ranking-based RAG by 12.5\%. Notably, even compared to the proprietary state-of-the-art reranker GTE-7B \cite{gte}, our method (335M) still demonstrates a 3.4\% improvement. 
These consistent performance gains extend across TriviaQA, PopQA and FM2, validating our approach's effectiveness across diverse scenarios.

\par
Our main contributions include:

\begin{itemize}

\item We introduce a novel metric called \textbf{Document Information Gain (DIG)}, to quantify each retrieved document's impact on the LLM's generation confidence. Different from semantic similarity, DIG can more accurately evaluate whether the document is helpful for generating a correct answer;

\item We develop a multi-task training strategy, which is used to optimize one reranker added after the retriever, with the aim of fitting the DIG score for each document. This strategy is designed from the exact distinguishing and accurate sorting perspectives, so as to filter out the irrelevant and select the most valuable documents for answer generation.

\item Integrating the DIG and the multi-task reranker, we propose \textbf{InfoGain-RAG}, a comprehensive framework for enhancing RAG. This framework can improve the quality of generation with both single and multiple retrievers, showing strong adaptability across vairous real-world settings with only an efficient, plug-and-play reranking module.
\end{itemize}

\section{Related Work} 

RAG has emerged as a promising solution to address fundamental limitations of LLMs. However, a key challenge in RAG systems lies in effectively evaluating and selecting the most valuable documents for answer generation. Existing document selections in RAG broadly follow three approaches:

The first approach optimizes retrievers through training on task-specific datasets. 
% RePlug \cite{shi2023replug} uses the black-box LLM's outputs as supervision signals to train the retriever, to prefer documents that reduce language model perplexity. 
RePlug \cite{shi2023replug} proposed a training pipeline that uses black-box LLM outputs as supervision signals to optimize the retriever, aiming to reduce LLM perplexity.
RA-DIT \cite{radit} proposed a dual instruction tuning framework that jointly optimizes both the LLM and retriever. Though useful, they struggle with multiple retrievers.

The second approach aims to evaluate retrieved documents utility by LLMs's self-reflection capabilities \cite{selfrag,corrective-rag}. Self-RAG introduces reflection tokens that allow the LLM to adaptively retrieve passages on-demand and critique both the retrieved content and its own generations. While effective in identifying valuable documents, multiple LLM calls introduce substantial computation overhead.

The third approach incorporates a reranker to reorder retrieved documents, typically including the open-source reranker BGE \cite{bge-reranker} and proprietary GTE-7B \cite{gte}. 
BGE is a small encoder initially trained on over 300M text pairs, then supervised fine-tuning on high-quality labeled data, while GTE-7B trains a large long-context LLM to learn the hybrid document representations (both dense and sparse). 
However, BGE is mainly trained to capture fine-grained semantic relationships, which may fail to select truly helpful documents, and GTE is computationally expensive for practical deployment.

\section{Method}

\begin{figure*}[ht!]
  \centering  
  \includegraphics[width=0.85\linewidth]{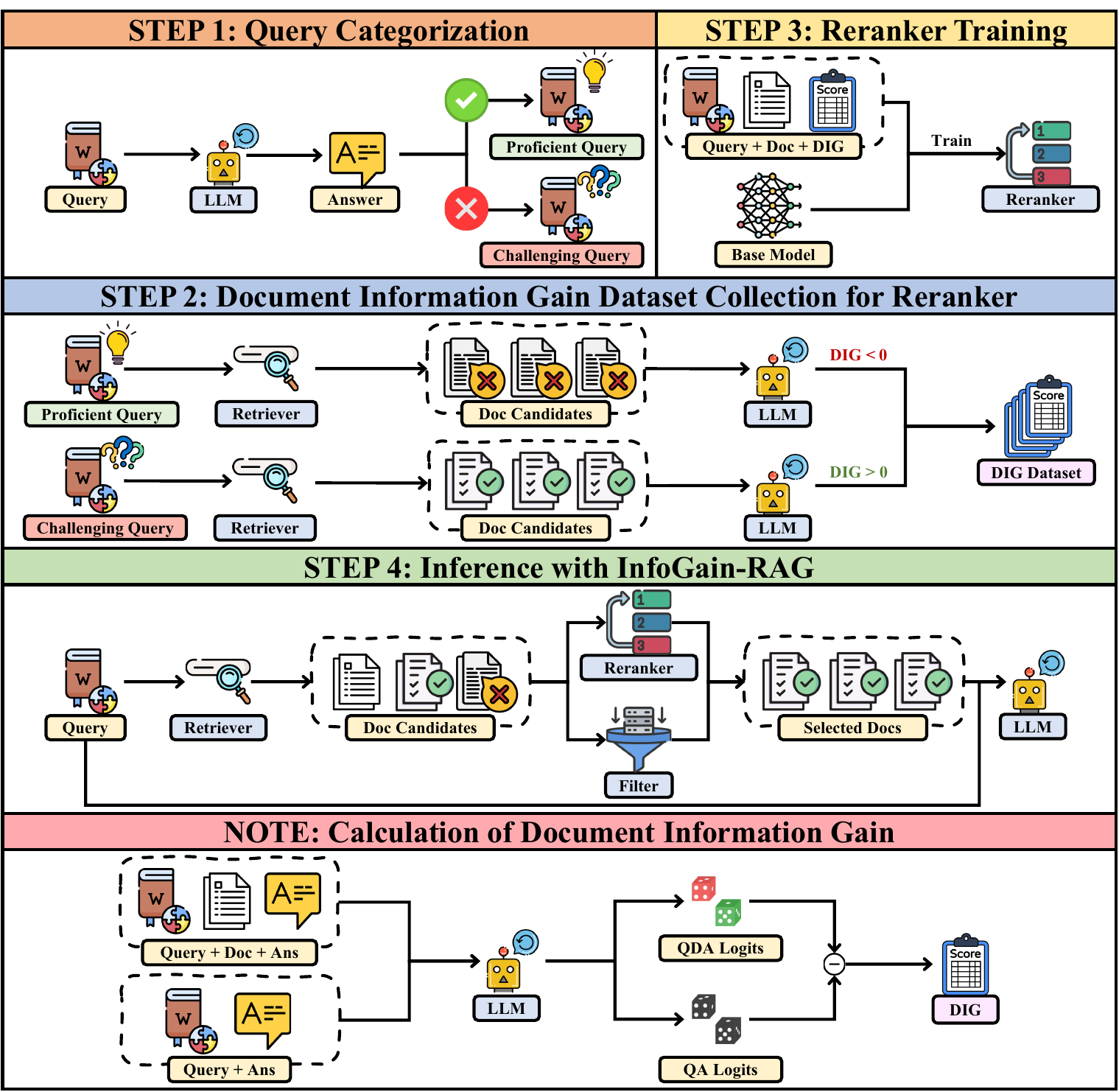}
  \caption{Illustrations of InfoGain-RAG. \textbf{STEP 1}: Distinguish proficient queries from challenging ones; \textbf{STEP 2}: Retrieve top-k documents for each query and calculate their DIG scores; \textbf{STEP 3}: Train the multi-task reranker; \textbf{STEP 4}: Inference with InfoGain-RAG; \textbf{NOTE}: Calculation of DIG.}
  \label{fig:method}
\end{figure*}

\par
In this section, we present InfoGain-RAG to address the key challenges discussed earlier. Our framework consists of two main components: (1) Document Information Gain (DIG), a metric that quantifies a document's contribution to correct answer generation by measuring changes in LLM's generation confidence scores, along with an efficient pipeline for collecting high-quality training data, and (2) a multi-task reranker that combines document relevance classification and ranking objectives to optimize document selection. By incorporating these, our framework enables effective document selection without requiring multiple LLM calls, making it both computationally efficient and practical for real-world applications.

\subsection{Document Information Gain}
\par
The core of InfoGain-RAG lies in quantifying each document's contribution to correct answer generation through calculating the information gain of each retrieval. This section details our methodology for computing DIG and utilizing it to build high-quality training data. The complete data collection pipeline is presented in Algorithm~\ref{alg:dig}. To compute DIG, we first propose a robust approach for estimating LLM's generation confidence, and then use this estimation to measure the information gain provided by each document.

\begin{algorithm}[H]
\footnotesize
\caption{DIG Data Collection Pipeline}
\label{alg:dig}  % 添加标签以便引用
\begin{algorithmic}[1]
\REQUIRE Query set $\mathcal{Q}$, Document corpus $\mathcal{D}$, LLM $\phi$
\ENSURE DIG dataset $\mathcal{T}$
\STATE $\mathcal{T} \gets \emptyset$
\FOR{each query $\mathrm{x} \in \mathcal{Q}$}
    \STATE Retrieve candidate documents $\mathrm{D_x}$ from $\mathcal{D}$
    \STATE Get confidence $p_{\phi}(\mathrm{y}|\mathrm{x})$ (defined in equation~\ref{eq:p_final}) without documents
    \FOR{each doc $\mathrm{d} \in \mathrm{D_x}$}
        \STATE Get confidence $p_{\phi}(\mathrm{y}|\mathrm{x},\mathrm{d})$ with document
        \STATE Calculate DIG (defined in equation~\ref{eq: dig})
        % $\text{DIG}(\mathrm{d}|\mathrm{x}) = p_{\phi}(\mathrm{y}|\mathrm{x},\mathrm{d}) - p_{\phi}(\mathrm{y}|\mathrm{x})$
        \STATE $\mathcal{T} \gets \mathcal{T} \cup \{(\mathrm{x},\mathrm{d},\text{DIG}(\mathrm{d}|\mathrm{x}))\}$
    \ENDFOR
\ENDFOR
\RETURN $\mathcal{T}$
\end{algorithmic}
\end{algorithm}

\subsubsection{Answer Generation Probability}
A key challenge in computing DIG is estimating the probability of a specific answer. A straightforward way would be to multiply the probabilities of individual tokens as the final confidence score. However, this approach faces two key challenges: First, it suffers from the length bias problem~\citep{Shi2021lengthbias} where longer sequences tend to receive lower scores as any single low token probability severely impacts the overall score. 
Second, treating all tokens equally fails to capture the strongest signal for generation quality~\cite{india-first} which initial tokens often provide. 
To address these, we propose a two-component approach:

\textbf{Sliding Window Smoothing}: To mitigate the length bias problem, we implement a sliding window smoothing mechanism. For each token $t_i$ in the answer sequence, its smoothed probability is calculated as:

\begin{equation}
\footnotesize
p_{\text{smooth}}(t_i) = \frac{1}{W}\sum_{j=i-\lfloor W/2 \rfloor}^{i+\lfloor W/2 \rfloor}{p(t_j)}
\label{eq:p_smooth}
\end{equation}
where $W$ is the window size and $p(t_j)$ represents the original token probability, obtained by normalizing LLM logits~\cite{logits}.

\textbf{Token Importance Weighting}: It is reported that initial tokens often carry stronger signals in model generation\cite{india-first}. Incorporating this observation, we apply higher weights to the first $k$ tokens when computing probability scores, as they typically contain core semantic information for the response.
% This aligns with findings from \cite{india-first} demonstrating the predictive power of early token probabilities.
The final formula is as follows:
\begin{equation}
\footnotesize
\begin{aligned}
p_\mathrm{\phi}(\mathrm{y}|\mathrm{x}) = 
\prod_{i=1}^{k}{(p_{\text{smooth}}(t_i))^{\omega_i \cdot \alpha}} \cdot 
\prod_{j=k+1}^{|\mathrm{y}|}{(p_{\text{smooth}}(t_j))}^{1-\alpha}
\label{eq:p_final}
\end{aligned}
\end{equation}
% \begin{equation}
% \footnotesize
% \begin{aligned}
% \log p_\mathrm{\phi}(\mathrm{y}|\mathrm{x}) = \alpha \sum_{i=1}^{k} \omega_i \cdot \log p_{\text{smooth}}(t_i) \\
% + (1-\alpha) \sum_{j=k+1}^{|\mathrm{y}|} {\log p_{\text{smooth}}(t_j)}
% \end{aligned}
% \end{equation}
where $\omega_i$ are the importance weights for the first $k$ tokens, $\alpha$ is a weight hyper-parameter, and $|\mathrm{y}|$ is the answer length.

\subsubsection{Calculation of DIG}
With a reliable approach to estimate answer generation probability, we now define the calculation of DIG, as shown in Figure~\ref{fig:method}(NOTE part). Unlike traditional relevance metrics that rely on lexical overlap or semantic similarity, DIG directly measures how much a document improves the LLM's confidence in generating the correct answer.

Formally, given an LLM $\mathrm{\phi}$, a query $\mathrm{x}$, and its corresponding ground truth answer $\mathrm{y}$, the DIG for a document retrieved $\mathrm{d}_i(\mathrm{d}_i \in \mathcal{D}, \mathcal{D}=\{\mathrm{d}_1,\mathrm{d}_2,\ldots,\mathrm{d}_{|\mathcal{D}|}\})$ is defined as:

\begin{equation}
\footnotesize
\mathrm{DIG}(\mathrm{d}_i|\mathrm{x}) \overset{def}{=} p_\mathrm{\phi}(\mathrm{y}|\mathrm{x},\mathrm{d}_i) - p_\mathrm{\phi}(\mathrm{y}|\mathrm{x})
% \mathrm{DIG}(\mathrm{d}_i|\mathrm{x}) = \log\frac{p_\mathrm{\phi}(\mathrm{y}|\mathrm{x})}{p_\mathrm{\phi}(\mathrm{y}|\mathrm{x},\mathrm{d}_i)}
\label{eq: dig}
\end{equation}
where $p_\mathrm{\phi}(\mathrm{y}|\mathrm{x},\mathrm{d}_i)$ represents the model's output confidence with both the query and the document, and $p_\mathrm{\phi}(\mathrm{y}|\mathrm{x})$ is the query-only confidence.

Based on above, we establish a data collection pipeline that begins by categorizing queries based on the model's baseline performance without retrieved documents, shown in Figure~\ref{fig:method} (STEP 1):

\begin{itemize}
\item \textbf{Model-Proficient Queries}: Queries that the LLM can answer correctly using only its inherent knowledge (i.e., high $p_\mathrm{\phi}(\mathrm{y}|\mathrm{x})$). These queries are particularly effective for identifying noisy documents through $\mathrm{DIG} < 0$, while positive DIG samples are naturally rare since external correct information adds little value to already-known answers.

\item \textbf{Model-Challenging Queries}: Queries that the LLM shows low confidence without external information (i.e., low $p_\mathrm{\phi}(\mathrm{y}|\mathrm{x})$). These queries facilitate us to identify helpful documents, as confidence increases ($\mathrm{DIG} > 0$).
\end{itemize}

% The DIG computation process involves two parallel inference paths:
% \begin{itemize}
% \item A baseline path processing the query $\mathrm{x}$ independently
% \item A document-augmented path combining eac  h document $\mathrm{d}_i$ with the query
% \end{itemize}

Based on DIG, documents are categorized into three groups (see Figure \ref{case-dig}):

\begin{itemize}
\item \textbf{DIG $>$ 0}: Documents that enhance the model's confidence, containing relevant and helpful information that should be prioritized during reranking.

\item \textbf{DIG $\approx$ 0}: Documents that neither improve nor diminish confidence and occur in two scenarios: (1) the document contains no meaningful information for answering the query, or (2) LLM has already mastered the required knowledge during pre-training, making additional correct information unnecessary.

\item \textbf{DIG $<$ 0}: Documents that reduce confidence and contain misleading or contradictory information that should be filtered out.
\end{itemize}

This categorization offers two key advantages: 1) quantitative measurement of document utility through DIG scores, enabling both automatic identification of high-quality documents and precisely filtering noise; and 2) fine-grained document prioritization through continuous DIG scores, which allows optimal document ordering during inference.

By computing DIG across diverse query-document pairs, we create a rich training dataset capturing both absolute relevance and relative importance of documents. This dataset serves as the foundation for training our specialized reranker, as detailed in the following section.

\subsection{Multi-task Reranker}

Building on DIG-scored training data collected above, we propose a multi-task learning strategy to train our reranker to select the most valuable documents for correct answer generation. The training objective combines Cross-Entropy (CE) loss and Margin loss to filter out noisy content and prioritize highly effective documents based on DIG scores. CE loss enables the model to distinguish between helpful and noisy documents through binary classification, while margin loss optimizes document ordering based on their DIG values. This unified training approach enables our reranker to simultaneously learn discriminative document classification and fine-grained ranking preferences, leading to robust document selection for RAG.

\subsubsection{Document Relevance Classification}

The first task focuses on the relevance determination of the retrievals through binary classification. Building upon the former collected data, we train the reranker to distinguish documents that have substantial contributions or potential harm to answer generation. Specifically, we employ CE loss to optimize the reranker $\theta$ to achieve this objective:

\begin{equation}
\footnotesize
\begin{aligned}
\min_{\theta} \quad L_{\text{CE}} = & \frac{1}{N} \sum_{i=1}^N \Big[-y_i \log(p(\mathrm{x}_i,\mathrm{d}_i)) \\
&\quad\quad\quad -(1-y_i)\log(1-p(\mathrm{x}_i,\mathrm{d}_i))\Big] \\
\text{s.t.} \quad & p(\mathrm{x}_i,\mathrm{d}_i) \in [0,1], y_i \in \{0,1\},\forall i=1,\ldots,N
\end{aligned}
\end{equation}

Here, $p(\mathrm{x_i,d_i})$ represents the predicted probability that document $\mathrm{d}_i$ will achieve a positive DIG score for query $\mathrm{x}_i$. The label $\mathrm{y}_i$ is determined by our previously computed DIG scores, with $\mathrm{y}_i=1$ for documents whose score is above upper decision boundary $b_1$ and $\mathrm{y}_i=0$ for those below lower decision boundary $b_2$. These thresholds effectively separate helpful documents from harmful ones. These hyper-parameters selection will be detailed in the experiment section. This classification-based learning not only helps identify useful documents but also facilitates better learning of relative document ordering through the joint training process.

\subsubsection{Document Ranking Optimization}

The second task focuses on learning relative document importance through pairwise comparison. Inspired by Circle Loss \cite{circle}, we introduce a margin-based learning objective that explicitly models the relative ordering of documents based on their DIG values. Given a query, this objective constrains the maximum score of negative query-document pairs to be lower than the minimum score of positive pairs:

\begin{equation}
\footnotesize
\begin{aligned}
\min_{\theta} \quad & L_{\text{Margin}} = \left[\max \left(s_n\right)-\min \left(s_p\right)\right]_{+} \\
\text{with} \quad & [x]_+ = \max(x,0)
\label{eq: equation total}
\end{aligned}
\end{equation}

where $s_n$ and $s_p$ denote scores for pairs with DIG values above $b_1$ and below $b_2$ respectively, and $\theta$ denotes reranker. To involve all samples in one process, we employ the LogSumExp function to approximate extremal value:
% \vspace{-1cm}
\begin{equation}
\footnotesize
\begin{aligned}
\max \left\{x_1, \ldots, x_n\right\} = & \log \left(\exp \left(\max \left(x_i\right)\right)\right) \approx LSE\left(x_n\right) ,
\\
\min \left\{x_1, \ldots, x_n\right\} = & - \max \left\{-x_1, \ldots, - x_n\right\} \\
& \approx -LSE\left(-x_n\right)
\label{eq: esti_min}
\end{aligned}
\end{equation}

where $LSE(x_n)$ is the LogSumExp function, with detailed derivation provided in Appendix \ref{Appendix B.1}.

Substitute the LogSumExp approximations into equation (\ref{eq: equation total}) and yield:
\begin{equation}
\footnotesize
\begin{aligned}
\min_{\theta} \quad L_{\text{Margin}} & \approx \left[L S E\left(\gamma\left(s_n\right)\right) - \left(-L S E\left(-\gamma\left(s_p\right)\right)\right)\right]_{+} \\
& \approx \log \left[1+\sum_{i=1}^K \sum_{j=1}^L \exp \left(\gamma\left(s_n^j-s_p^i\right)\right)\right]
\label{eq: esti_loss}
\end{aligned}
\end{equation}
where $\gamma$ is a scaling factor controlling the contribution of non-extremal pairs and $K$ and $L$ denote the number of positive and negative document pairs. 
Detailed derivation is provided in Appendix \ref{Appendix B.2}. 
Softplus is used to smooth the ReLU function:
\begin{equation}
\footnotesize
\operatorname{Softplus}(x)=\log \left(1+e^x\right) \approx [x]_{+}
\label{eq: softplus}
\end{equation}

By integrating CE loss and margin loss with weight $\beta$, our multi-task training objective enables the reranker to jointly optimize DIG and inter-document relationships: 

\begin{equation}
\footnotesize
L_{\text{total}} = \beta L_{\text{CE}} + (1-\beta) L_{\text{Margin}}
\label{eq: loss_total}
\end{equation}

\begin{figure}
    \centering
    \includegraphics[width=0.9\linewidth]{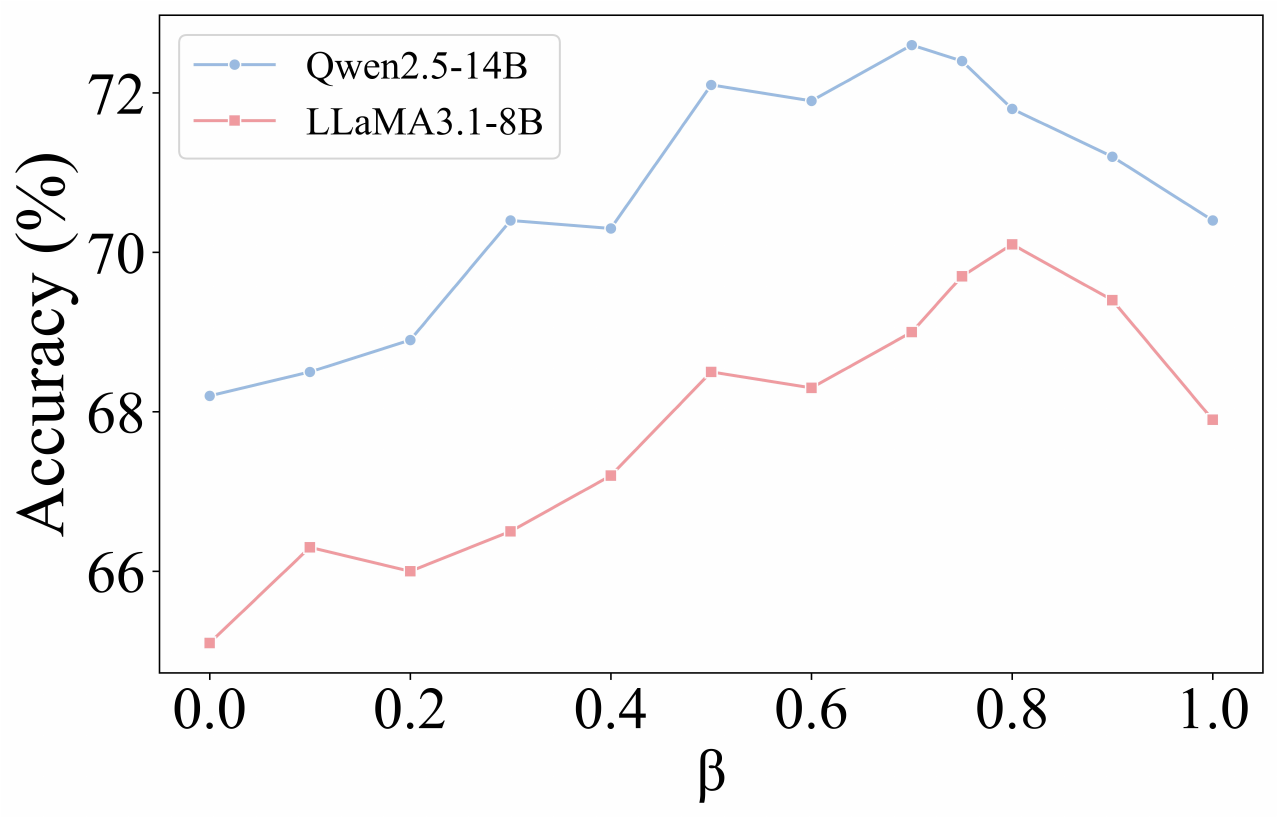}
    \caption{The relationship between the hyper-parameter $\beta$ and accuracy on TriviaQA, LLaMA3.1-8B achieves optimum at $\beta=0.8$, while Qwen2.5-14B at 0.7.}
    \label{fig:hyper-beta}
\end{figure}

This unified approach produces a robust reranker that considers both absolute document relevance and relative ordering preferences within the retrieved documents, leading to more effective document reranking and filtering for RAG systems (see Figure~\ref{fig:hyper-beta} for empirical study on balancing these two objectives via hyper-parameter $\beta$).

During inference, InfoGain-RAG enhances naive RAG pipelines by adding an efficient document reranking step while maintaining low computational overhead, as illustrated in Figure~\ref{fig:method} (STEP 4). The process begins with document retrieval, followed by our trained reranker which both reorders documents and filters out those below a quality threshold. The filtered and reranked documents are then passed to LLM for final answer generation while only calling once.

\section{Experiment}

\begin{table*}[htbp]
\small  
\setlength{\tabcolsep}{3pt}
    \centering
    \caption{Performance Comparison of RAG Reranking approaches with single-retriever (Contriever).}
    \label{tab:performance-comparison}
    \begin{adjustbox}{max width=\textwidth}
    \begin{tabular}{l|cccc|cccc|cccc|cccc}
    \toprule[1.5pt]
    \multirow{2}{*}{\textbf{Model}} & \multicolumn{4}{c|}{\textbf{TriviaQA}} & \multicolumn{4}{c|}{\textbf{NaturalQA}} & \multicolumn{4}{c|}{\textbf{PopQA}} & \multicolumn{4}{c}{\textbf{FM2}} \\
    & RAG & BGE(550M)$^\S$ & GTE(7B)$^\Diamond$ & Ours(355M) & RAG & BGE(550M)$^\S$ & GTE(7B)$^\Diamond$ & Ours(355M) & RAG & BGE(550M)$^\S$ & GTE(7B)$^\Diamond$ & Ours(355M) & RAG & BGE(550M)$^\S$ & GTE(7B)$^\Diamond$ & Ours(355M) \\
    \midrule[1pt]
    Qwen2.5-0.5B & 48.5\% & 48.6\% & 49.5\% & \textbf{55.8\%} & 22.5\% & 27.3\% & 29.5\% & \textbf{35.3\%} & 26.5\% & 35.7\% & 35.3\% & \textbf{36.5\%} & 53.0\% & 52.3\% & 55.6\% & \textbf{58.7\%} \\
    Qwen2.5-1.5B & 50.4\% & 59.1\% & 63.3\% & \textbf{66.3\%} & 30.7\% & 39.5\% & 45.2\% & \textbf{47.2\%} & 31.3\% & 41.3\% & \textbf{44.2\%} & 43.0\% & 69.1\% & 69.5\% & 71.1\% & \textbf{73.9\%} \\
    % Qwen2.5-3B & 51.7\% & 65.7\% & 65.8\% & \textbf{68.2\%} & 32.4\% & 38.6\% & 49.3\% & \textbf{50.9\%} & 31.9\% & 43.1\% & \textbf{47.7\%} & 47.3\% & 69.0\% & 72.1\% & 76.5\% & \textbf{79.2\%} \\
    Qwen2.5-7B & 52.9\% & 67.0\% & 69.5\% & \textbf{72.1\%} & 36.3\% & 41.8\% & 49.9\% & \textbf{53.6\%} & 32.4\% & 43.4\% & 43.7\% & \textbf{47.6\%} & 72.5\% & 74.5\% & 77.8\% & \textbf{79.9\%} \\
    Qwen2.5-14B & 56.1\% & 68.4\% & 71.1\% & \textbf{72.9\%} & 36.0\% & 42.7\% & 52.5\% & \textbf{53.8\%} & 31.8\% & 44.1\% & 45.9\% & \textbf{49.4\%} & 72.6\% & 75.7\% & 76.4\% & \textbf{79.4\%} \\
    Qwen2.5-32B & 58.7\% & 70.3\% & 72.0\% & \textbf{74.7\%} & 36.4\% & 42.1\% & 53.7\% & \textbf{55.9\%} & 32.3\% & 45.5\% & 48.1\% & \textbf{50.5\%} & 73.7\% & 75.6\% & 79.0\% & \textbf{81.2\%} \\
    Qwen2.5-72B & 59.9\% & 70.6\% & 73.4\% & \textbf{76.3\%} & 40.3\% & 44.9\% & 53.9\% & \textbf{58.1\%} & 34.0\% & 44.8\% & 49.5\% & \textbf{51.4\%} & 73.6\% & 75.9\% & 80.4\% & \textbf{83.4\%} \\
    Qwen3-8B & 57.9\% & 67.6\% & 71.1\% & \textbf{72.3\%} & 34.0\% & 41.5\% & 50.9\% & \textbf{52.6\%} & 32.1\% & 43.6\% & 46.5\% & \textbf{49.1\%} & 71.4\% & 76.1\% & \textbf{80.9\%} & 80.0\% \\
    \midrule[1pt]
    LLaMA3.1-8B & 55.1\% & 65.5\% & 67.5\% & \textbf{70.4\%} & 33.6\% & 39.4\% & 46.9\% & \textbf{50.7\%} & 31.7\% & 41.3\% & 44.6\% & \textbf{47.1\%} & 74.3\% & 77.6\% & 79.5\% & \textbf{81.2\%} \\
    LLaMA3.1-70B & 54.5\% & 67.9\% & 67.4\% & \textbf{71.3\%} & 35.1\% & 39.9\% & 48.6\% & \textbf{51.6\%} & 30.4\% & 43.0\% & 47.2\% & \textbf{47.6\%} & 77.0\% & 79.5\% & 81.1\% & \textbf{82.4\%} \\
    LLaMA3.1-405B & 56.7\% & 69.2\% & 73.8\% & \textbf{74.6\%} & 35.8\% & 41.5\% & 52.3\% & \textbf{53.3\%} & 30.5\% & 43.4\% & 47.3\% & \textbf{49.5\%} & 75.9\% & 77.6\% & 80.6\% & \textbf{83.1\%} \\
    \midrule[1pt]
    Gemma-2-9B & 54.3\% & 64.4\% & 69.0\% & \textbf{71.3\%} & 34.3\% & 39.6\% & 44.6\% & \textbf{56.6\%} & 31.4\% & 43.9\% & 45.5\% & \textbf{49.3\%} & 75.4\% & 78.5\% & 80.9\% & \textbf{81.5\%} \\
    Gemma-2-27B & 59.6\% & 68.5\% & 70.9\% & \textbf{74.3\%} & 37.6\% & 42.3\% & 51.5\% & \textbf{57.4\%} & 33.1\% & 45.4\% & 49.4\% & \textbf{50.3\%} & 76.3\% & 78.4\% & \textbf{82.1\%} & 81.6\% \\
    \midrule[1pt]
    DeepSeek-V3 & 56.0\% & 68.0\% & 72.0\% & \textbf{73.4\%} & 37.6\% & 42.5\% & 50.7\% & \textbf{55.1\%} & 30.8\% & 43.4\% & 48.6\% & \textbf{49.7\%} & 75.7\% & 77.5\% & 78.6\% & \textbf{80.2\%} \\
    DeepSeek-R1 & 60.4\% & 71.7\% & \textbf{75.7\%} & 75.2\% & 40.8\% & 44.8\% & 56.8\% & \textbf{58.8\%} & 31.2\% & 45.3\% & 51.1\% & \textbf{51.6\%} & 77.1\% & 78.9\% & 80.3\% & \textbf{83.8\%} \\
    \midrule[1pt]
    Claude-Sonnet$^\dagger$ & 54.5\% & 68.4\% & 70.7\% & \textbf{73.9\%} & 36.7\% & 41.1\% & 52.4\% & \textbf{55.2\%} & 31.6\% & 43.1\% & 48.9\% & \textbf{50.4\%} & 76.0\% & 78.4\% & \textbf{80.9\%} & 80.8\% \\
    ChatGPT$^\ddagger$ & 62.0\% & 69.0\% & 72.1\% & \textbf{74.1\%} & 37.1\% & 42.7\% & \textbf{55.9\%} & 54.5\% & 32.0\% & 43.5\% & 48.0\% & \textbf{48.5\%} & 71.9\% & 73.2\% & 75.0\% & \textbf{75.3\%} \\
    GPT-4o$^\star$ & 57.2\% & 69.2\% & 74.4\% & \textbf{75.4\%} & 37.5\% & 41.6\% & 53.1\% & \textbf{57.2\%} & 31.4\% & 43.3\% & 48.3\% & \textbf{49.2\%} & 76.6\% & 75.1\% & 78.8\% & \textbf{82.2\%} \\
    GPT-4.1$^\S$ & 58.6\% & 70.8\% & 76.1\% & \textbf{76.4\%} & 35.1\% & 41.7\% & 55.6\% & \textbf{56.2\%} & 30.9\% & 45.4\% & \textbf{51.3\%} & 50.4\% & 75.2\% & 77.1\% & 76.4\% & \textbf{80.4\%} \\
    \bottomrule[1.5pt]
    \end{tabular}
    \end{adjustbox}
    \vspace{1mm}
    \raggedright
    \footnotesize{$^\S$BGE-Reranker-Large (550M). $^\Diamond$ Proprietary GTE-Reranker (7B). $^\dagger$241022 version. $^\ddagger$240125 version. $^\star$241120 version. $^\S$ 20250414 version.}
\end{table*}

We evaluate InfoGain-RAG in four experiment series. First, we compare it with the open-source reranker of BGE-Reranker-Large~\cite{bge-reranker} trained on 300M samples, and the state-of-the-art proprietary reranker GTE-7B~\cite{gte}. Second, we compare with retriever optimization approaches like RePlug~\cite{shi2023replug} and RADIT~\cite{radit}, and self-reflection approaches like Self-RAG~\cite{selfrag} and CRAG~\cite{corrective-rag}. Third, we test InfoGain-RAG on combined documents retrieved from Contriever~\cite{contriever}, BM25~\cite{bm25} and DPR~\cite{DPR} to demonstrate its capability to handle multiple retrievers. Last, several ablation studies are conducted to verify the effectiveness from different aspects. The datasets and models we used are publicly accessible.

\subsection{Setup}
\paragraph{Tasks and Datasets.}
We experiment on two tasks of four English datasets: (1) \textbf{open-domain question answering}, including TriviaQA \cite{joshi2017triviaqa}, NaturalQA \cite{kwiatkowski2019natural}, and PopQA \cite{popqa}; (2) \textbf{fact verification}, FM2 \cite{fm2}. 
We use the December 2018 Wikipedia dump \cite{DPR} as the retrieval corpus.

\paragraph{Models and Metrics.} 
All evaluations are conducted across both proprietary LLMs (GPT-4o-20241120, ChatGPT-20240125, and Claude-3.5-Sonnet-20241022) and open-source models (LLaMA3.1, Qwen2, Gemma2, DeepSeek-V3, and DeepSeek-R1). We adpot Exact Match (EM) accuracy~\cite{rajpurkar-etal-2016-squad} as the metric. EM provides a strict evaluation of response accuracy while accommodating multiple correct answer formats, as it compares the model outputs with all valid answers provided.

\paragraph{Implementation Details.}
We sample 110K queries from TriviaQA dataset (with train-test overlap removed) and calculate DIG scores for all collected \textit{<query, answer, document>}  triplets using Qwen2.5-7B. The scoring results in three categories: 70K triplets with high positive gain (>$b_1=0.5$), 150K triplets with negative gain (<$b_2=-0.2$), and 1200K triplets showing negligible information gain ($-0.05 \sim 0.05$). From these scored triplets, we create a unified training dataset of 88K samples through different sampling strategies for each loss: for CE loss, we sample balanced query-document pairs with equal numbers of positive and negative samples (68K), while for margin loss, we sample query-document groups (34K) where each query is paired with 3-5 high-DIG documents and augmented with additional negative and negligible documents. 

For experimental settings, we implement our reranker using RoBERTa-large \cite{liu2019roberta} to rerank the top 100 documents retrieved by Contriever~\cite{contriever}. 
Our reranker is trained on an A800 GPU using Adam optimizer with a learning rate of 5e-6 and $\beta$ value of 0.75. 
For DIG calculation, we set importance weights $\omega_i$ to 0.8 for the first $k=3$ tokens and use $\alpha=0.6$ for balancing token probabilities. 
During inference, we select the top 4 documents and employ a document filtering threshold of 0.2 while retaining all candidates that exceed this threshold. This threshold is slight different from $b_1$, as the the addition of margin loss would widen the score distribution of valid samples. Notably, to ensure minimal context for generation, we retain at least 2 documents if fewer exceed the filtering threshold.

\subsection{Results}

We first present InfoGain-RAG's performance with single retriever across different LLMs and benchmarks, comparing it with naive RAG and reranking approaches. We then show its effectiveness in multiple retriever settings. Finally, we demonstrate our method's advantages over self-reflection and retriever-optimization approaches.

\paragraph{Comparison to Reranking approaches with Single Retriever.}
Table~\ref{tab:performance-comparison} compares InfoGain-RAG (355M) against naive RAG, BGE-Reranker (550M) and GTE-Reranker (7B, SOTA) across different models and datasets. As shown in the results, InfoGain-RAG substantially improves over naive RAG and BGE-Reranker, while surpassing the far larger GTE-Reranker in most cases. On TriviaQA, for instance, DeepSeek-V3 achieves 72.0\% with GTE-Reranker and 73.4\% with InfoGain-RAG, while Qwen2.5-72B reaches 76.3\% with InfoGain-RAG, surpassing naive RAG by 16.4\%, BGE-Reranker by 5.7\%, and GTE-Reranker by 2.9\%. Moveover, these improvements hold across both model scales and families - from smaller models like Qwen2.5-1.5B (+15.9\% over naive RAG) to larger ones like LLaMA3.1-405B (+17.9\%).

Trained on TriviaQA, InfoGain-RAG demonstrates strong generalization ability across different datasets and tasks. It improves Qwen2.5-72B's accuracy on NaturalQA by 17.8\% and PopQA by 17.4\% over naive RAG, with particularly notable gains on FM2 from 73.6\% to 83.4\%. 

In particular, our reranker achieves these results with just 88K training samples and merely 335M parameters, compared to BGE-Reranker's 300M samples and GTE-Reranker's 7B parameters(see Appendix \ref{Appendix C} for comparisons with the GTE family).

\begin{figure}
    \centering
    \includegraphics[width=0.9\linewidth]{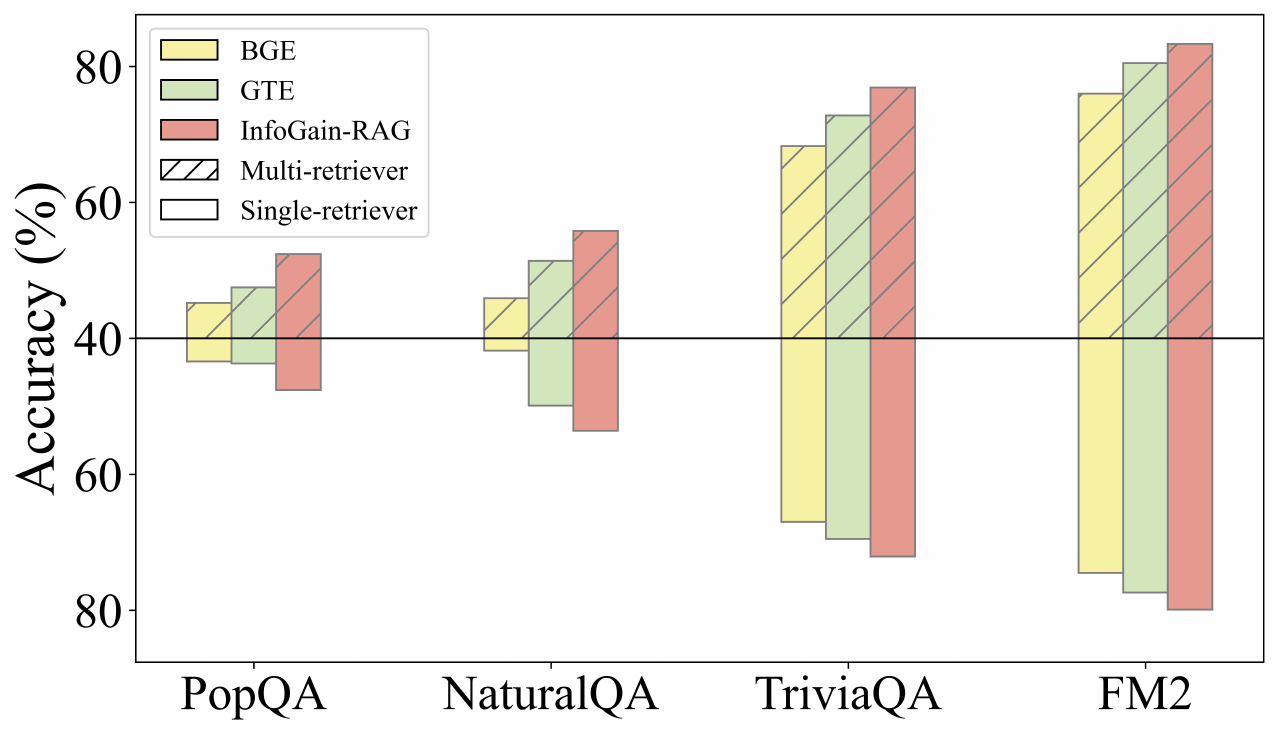}
    \caption{Performance comparison of Qwen2.5-7B across different datasets with single retriever and multiple retrievers.}
    \label{multi_retriever}
\end{figure}

\paragraph{Comparison to Reranking approaches with Multiple Retrievers.}
 
InfoGain-RAG maintains consistent superiority with multiple retrievers.

As shown in Figure \ref{multi_retriever}, our reranker achieves the best performance on all four tasks. Specifically, it improves by 9.9\% over BGE-Reranker on NaturalQA and by 4.9\% over GTE-Reranker on PopQA. Additionally, we observe that all rerankers show improvements in the multi-retriever setting compared to the single-retriever setting. Notably, our method achieves the largest performance gains (when comparing multi-retriever to single-retriever settings) on most tasks, with an average improvement of 3.8\%. This clearly demonstrates the superior effectiveness of our reranker in multi-retriever scenarios.

\begin{figure}[!htbp]
\centering
  \includegraphics[width=\linewidth]{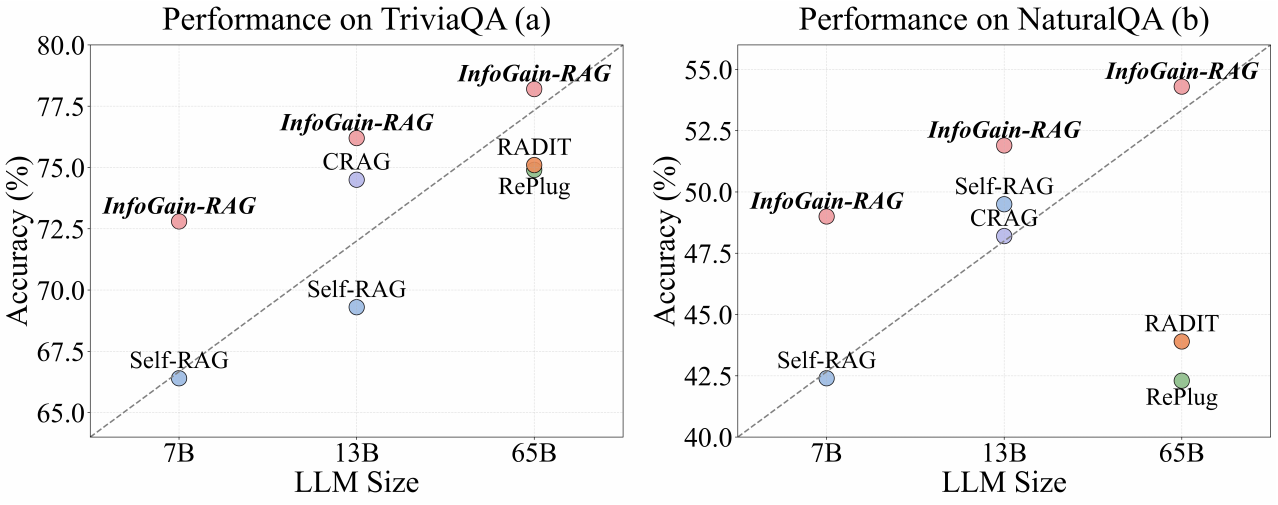}
  \caption {Performance Comparison with self-reflection (7B, 13B) and retriever-optimization (65B) approaches on TriviaQA (a) and NaturalQA (b). We strictly followed the experimental settings of each baseline approach for fair comparison.}
  \label{fig:sota-comparison}
\end{figure}

\paragraph{Comparison with Self-Reflection and Retriever-Optimization approaches.}
As shown in Figure~\ref{fig:sota-comparison}, we evaluate InfoGain-RAG against two types of RAG approaches.
For self-reflection, our approach outperforms both Self-RAG\cite{selfrag} and CRAG\cite{corrective-rag}. 
With LLaMA2-13B as the base model, InfoGain-RAG achieves 76.2\% accuracy on TriviaQA and 51.9\% on NaturalQA, surpassing Self-RAG (69.3\%, 49.5\%) and CRAG (74.5\%, 48.2\%) while avoiding multiple LLM inference calls.
For retriever-optimization, InfoGain-RAG shows substantial improvements using LLaMA-65B, reaching 78.2\% on TriviaQA and 54.3\% on NaturalQA. This outperforms both RePlug\cite{shi2023replug} (74.9\%, 42.3\%) and RADIT\cite{radit} (75.1\%, 43.9\%). 

\subsection{Ablation Study}
In this section, we conduct comprehensive ablation studies to systematically evaluate the critical components across InfoGain-RAG: 1) examining whether using different base models to generate DIG data will affect the final effect, 2) verifying whether the multi-task learning strategy can bring greater improvement compared to each individual task, and 3) assessing the impact of document filtering during inference.
\paragraph{LLM-agnostic DIG-data Collection.}
Table \ref{DIG-calculation} demonstrates that InfoGain-RAG's performance remains consistent regardless of which LLM is used for DIG data collection. Despite the changes in the DIG scores of each model due to factors such as structure and size, the trained reranker achieves similar accuracy on TriviaQA. This performance shows that InfoGain-RAG can identify the intrinsic query-document correlations independent of the LLM used for data collection, validating its robustness as a general framework.

\begin{table}[ht]
\footnotesize
\caption{Compared results of rerankers trained using DIG scores from different base LLMs on  TriviaQA.}
\label{DIG-calculation}
\begin{tabular*}{\linewidth}{@{\extracolsep{\fill}}l|ccc}
\toprule[1.5pt]
\makecell[c]{\textbf{Model}} & \makecell[c]{\textbf{RAG}} & \makecell[c]{\textbf{Ours} \\ \textbf{(DIG-Qwen)}} & \makecell[c]{\textbf{Ours} \\ \textbf{(DIG-LLaMA)}} \\
\midrule[1pt]
Qwen2.5-7B & 52.9\% & \textbf{72.1\%} & 68.8\% \\
Qwen2.5-14B & 56.1\% & 72.9\% & \textbf{74.2\%} \\
Qwen2.5-72B & 59.9\% & \textbf{76.3\%} & 75.0\% \\
LLaMA3.1-8B & 55.1\% & 70.4\% & \textbf{72.1\%} \\
LLaMA3.1-70B & 54.5\% & \textbf{71.3\%} & 70.2\% \\
LLaMA3.1-405B & 56.7\% & \textbf{74.6\%} & 73.0\% \\
\bottomrule[1.5pt]
\end{tabular*}
\end{table}

\paragraph{Single or Multi-task Reranker Training.}
Table \ref{loss} compares the performance differences of single CE or Margin task to the multi-task training. We can see that the combined strategy consistently outperforms individual loss across two types of models. For example, Qwen2.5-72B can get an accuracy of 76.8\% with the multi-task training on TriviaQA, but only 73.0\% for CE and 71.4\% for margin loss. 
The large improvement demonstrates that the absolute relevance judgments can be combined with the relative rankings to achieve more robust document selection.

\begin{table}[ht]
\footnotesize
    \caption{Performance differences of single CE or Margin task to the multi-task training across models. The testings is conducted on TriviaQA.}
    \label{loss}
    \begin{adjustbox}{max width=\linewidth}
    \begin{tabular*}{\linewidth}{@{\extracolsep{\fill}}l|ccc}
    \toprule[1.5pt]
    \makecell[c]{\textbf{Model}}
    & \makecell[c]{\textbf{Ours} \\ \textbf{(CE loss)}} 
    & \makecell[c]{\textbf{Ours} \\ \textbf{(Margin loss)}} 
    & \makecell[c]{\textbf{Ours} \\ \textbf{(Multi-loss)}}
    \\
    \midrule[1pt]
    Qwen2.5-7B & 67.6\% & 68.2\% & \textbf{71.8\%}\\
    Qwen2.5-14B & 70.1\% & 67.9\% & \textbf{72.7\%}\\
    Qwen2.5-72B & 73.0\% & 71.4\% & \textbf{76.8\%}\\
    LLaMA3.1-8B & 68.2\% & 65.3\% & \textbf{70.7\%}\\
    LLaMA3.1-70B & 69.5\% & 67.1\% & \textbf{71.4\%}\\
    LLaMA3.1-405B & 73.6\% & 70.8\% & \textbf{74.2\%}\\
    \bottomrule[1.5pt]
    \end{tabular*}
    \end{adjustbox}
\end{table}

\paragraph{Document Filtering during Inference.} 
In table~\ref{filter} we test the effectiveness of document filtering during inference with the threshold of 0.2. 
Here, non-filtering means all retrieved documents are ranked without being filtered. 
It can be observed that peformances are better with filtering than non-filtering. 
For instance, Qwen2.5-72B improves from 73.6\% to 76.8\%, and LLaMA3.1-405B gains from 71.2\% to 74.6\%. 
These observations jointly confirm that identifying and removing potentially noisy contents is beneficial for final performance.

\begin{table}[ht]
\footnotesize
    \caption{Performance validations of retrieved document filtering operations. All results are tested on TriviaQA.}
    \label{filter}
    \begin{adjustbox}{max width=\linewidth}
    \begin{tabular*}{\linewidth}{@{\extracolsep{\fill}}l|ccc}
    \toprule[1.5pt]
    \makecell[c]{\textbf{Model}} & \textbf{RAG} & \makecell[c]{\textbf{Ours} \\ \textbf{(Non-filtering)}} & \makecell[c]{\textbf{Ours} \\ \textbf{(Filtering)}}\\
    \midrule[1pt]
    Qwen2.5-7B & 52.9\% & 68.2\% & \textbf{71.8\%} \\
    Qwen2.5-14B & 56.1\% & 71.8\% & \textbf{72.9\%} \\
    Qwen2.5-72B & 59.9\% & 73.6\% & \textbf{76.3\%} \\
    LLaMA3.1-8B & 55.1\% & 67.8\% & \textbf{70.4\%} \\
    LLaMA3.1-70B & 54.5\% & 68.2\% & \textbf{71.3\%} \\
    LLaMA3.1-405B & 56.7\% & 71.2\% & \textbf{74.6\%} \\
    \bottomrule[1.5pt]
    \end{tabular*}
    \end{adjustbox}
\end{table}

\section{Conclusion}
In this paper, we present a novel framework InfoGain-RAG to address the critical challenge of RAG about filtering out semantically misaligned and noisy retrieved content.
By introducing a principled DIG metric coupled with a multi-task reranker learning strategy, InfoGain-RAG effectively quantifies document utility and optimizes both filtering and reranking processes. Comprehensive experiments across proprietary and open-source LLMs demonstrate substantial improvements across multiple benchmarks while maintaining lower computational overhead compared to existing approaches. The effectiveness and economic applicability of the framework suggest the feasibility of InfoGain-RAG, as it can offer a reliable solution for RAG in partical application.

\section{Limitation}
While InfoGain-RAG demonstrates strong performance improvements, several limitations warrant discussion. The current implementation has only been tested on text modalities, though it is theoretically extensible to other modalities such as visual or code data. Computational constraints limit the reranker to 335M parameters rather than larger models (7B+), which could offer better performance but may significantly increase inference latency in practical applications. Additionally, the DIG metric, while effective, cannot distinguish factual inaccuracies in retrieved documents, which may require an extra module to address this issue. We hope more efforts can be devoted to addressing these limitations collaboratively in the future.

\bibliography{custom}

\newpage
\appendix

\newpage
\onecolumn
\section{DIG Cases}
\label{Appendix A}

% \FloatBarrier
\begin{figure}[H]
  \centering
  \includegraphics[width=0.85\textwidth]{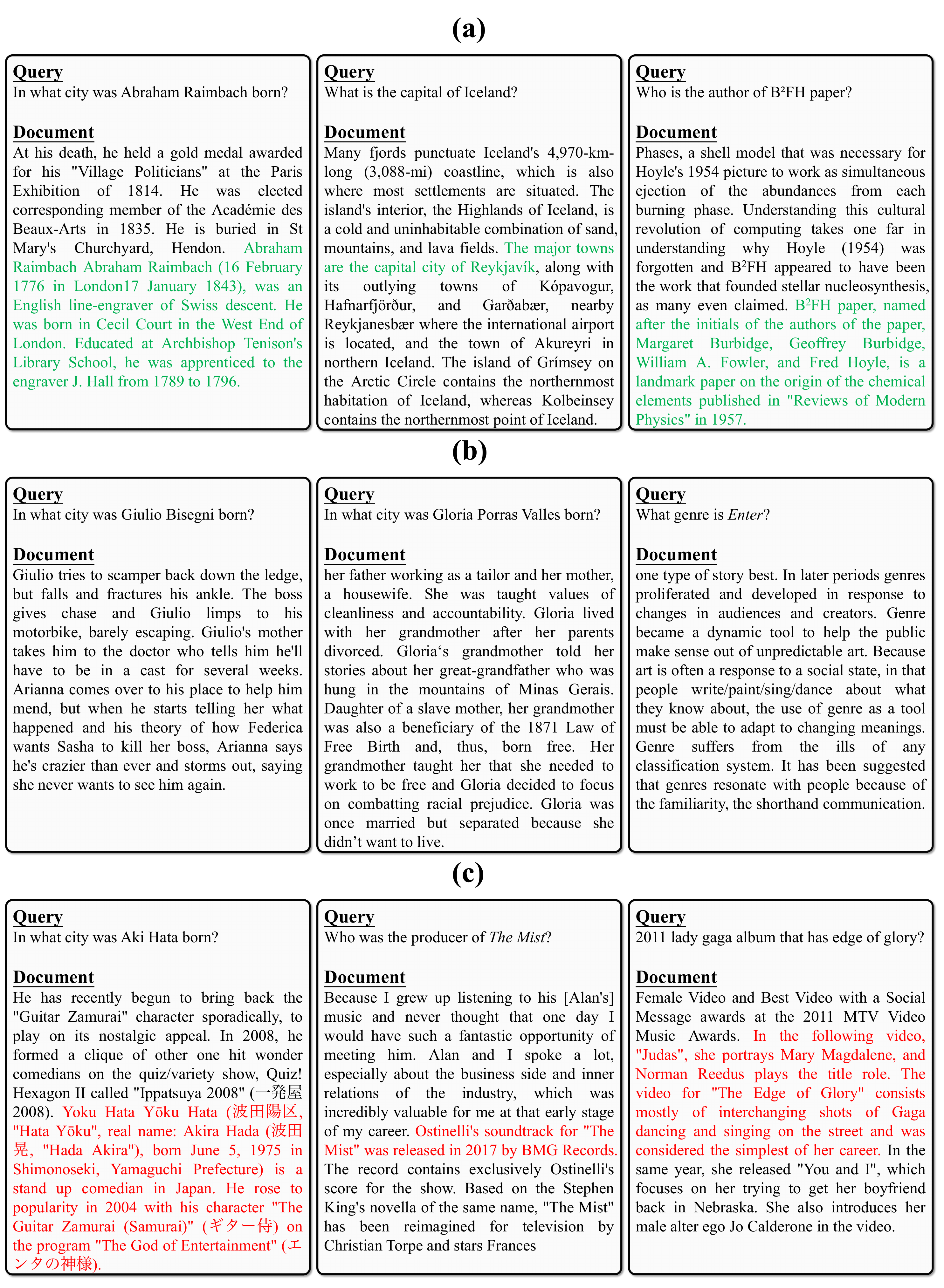}
  \caption {Retrieved documents of which DIG > 0 (a), DIG $\approx$ 0 (b), and DIG < 0 (c) for the given query.}
    \label{case-dig}
\end{figure}

\newpage
\begin{figure}[!htbp]
  \centering
  \includegraphics[width=0.85\textwidth]{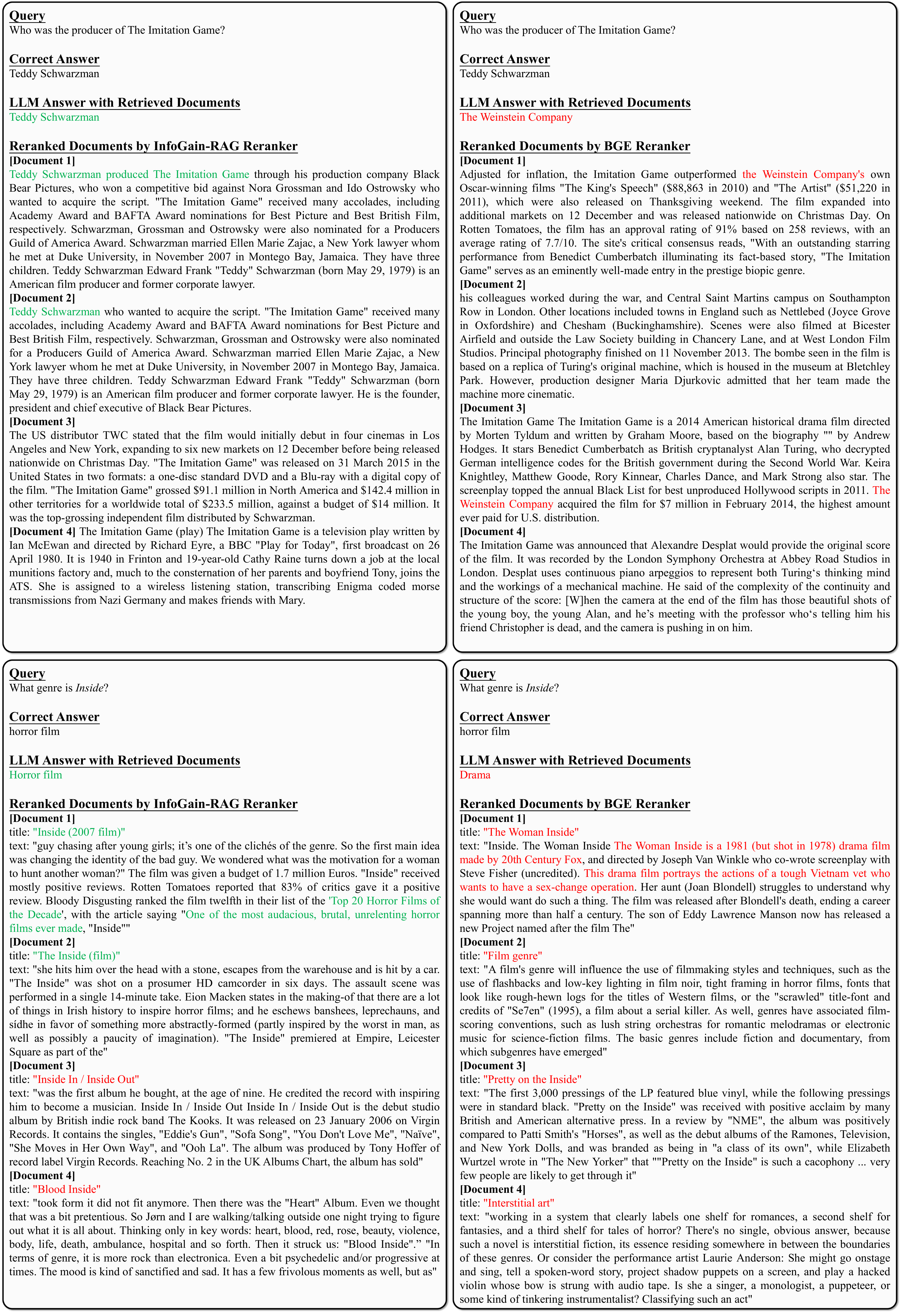}
  \caption {Comparison of documents retrieved by InfoGain-RAG reranker and BGE reranker.}
    \label{case-reranker}
\end{figure}

\twocolumn
\section{Mathematical Derivations}
\label{Appendix B}
In this section, we provide detailed mathematical derivations for two key components of margin loss: (1) how LogSumExp (LSE) function approximates the maximum function, and (2) the complete derivation steps of our margin loss formulation based on LSE.
\subsection{LSE Approximation of Maximum Function}
\label{Appendix B.1}
The LogSumExp function is defined as:
\begin{equation}
\footnotesize
LSE(x_1,\ldots,x_n) = \log(\sum_{i=1}^n \exp(x_i))
\end{equation}
First, we prove that LSE provides an upper bound for the maximum function. For any $i$:
\begin{equation}
\footnotesize
\begin{aligned}
LSE(x_1, \ldots, x_n) &= \log\left(\sum_{j=1}^n \exp(x_j)\right) \\
&\geq \log(\exp(x_i)) \\
&= x_i
\end{aligned}
\end{equation}

Since this holds for all $i$, we have:
\begin{equation}
\footnotesize
LSE(x_1,\ldots,x_n) \geq \max(x_1,\ldots,x_n)
\end{equation}
Let $x^* = \max(x_1,\ldots,x_n)$. We can rewrite LSE as:
\begin{equation}
\footnotesize
\begin{aligned}
&LSE(x_1,\ldots,x_n) = \log\left(\sum_{i=1}^n \exp(x_i)\right) \\
&= \log\left(\exp(x^*)\sum_{i=1}^n \exp(x_i-x^*)\right) \\
&= x^* + \quad \log\left(1 + \sum_{i:x_i\neq x^*} \exp(x_i-x^*)\right)
\end{aligned}
\end{equation}

Since $x_i-x^* \leq 0$ for all $i$ (with equality only when $x_i=x^*$), and typically $x_i-x^* \ll 0$ for $x_i \neq x^*$, we have:
\begin{equation}
\footnotesize
\exp(x_i-x^*) \rightarrow 0 \text{ when } x_i-x^* \ll 0
\end{equation}

Therefore:
\begin{equation}
\footnotesize
\log(1 + \sum_{i:x_i\neq x^*} \exp(x_i-x^*)) \rightarrow 0
\end{equation}

This yields our final approximation:
\begin{equation}
\footnotesize
LSE(x_1,\ldots,x_n) \approx x^* = \max(x_1,\ldots,x_n)
\end{equation}

The approximation becomes more accurate as the differences between the maximum value and other values increase.
\subsection{Derivation of Margin Loss}
\label{Appendix B.2}
Starting from the initial margin loss formulation:
\begin{equation}
\footnotesize
L_{\text{Margin}} \approx \left[L S E\left(\gamma\left(s_n\right)\right) - \left(-L S E\left(-\gamma\left(s_p\right)\right)\right)\right]_{+}
\end{equation}
We can expand this expression:

\begin{equation}
\footnotesize
\begin{aligned}
& \left[L S E\left(\gamma\left(s_n\right)\right) - \left(-N L S E\left(\gamma\left(s_p\right)\right)\right)\right]_{+} \\
& = \left[\log \sum_{j=1}^L \exp \left(\gamma\left(s_n^j\right)\right)+\log \sum_{i=1}^K \exp \left(\gamma\left(-s_p^i\right)\right)\right]_{+} \\
& = \left[\log \left(\sum_{j=1}^L \exp \left(\gamma\left(s_n^j\right)\right) \sum_{i=1}^K \exp \left(\gamma\left(-s_p^i\right)\right)\right)\right]_{+} \\
& = \left[\log \sum_{i=1}^K \sum_{j=1}^L \exp \left(\gamma\left(s_n^j-s_p^i\right)\right)\right]_{+}
\end{aligned}
\end{equation}

Finally, using the softplus function to smooth the ReLU operation:
\begin{equation}
\footnotesize
\begin{aligned}
L_{\text{Margin}} \approx \log \left[1+\sum_{i=1}^K \sum_{j=1}^L \exp \left(\gamma\left(s_n^j-s_p^i\right)\right)\right]
\end{aligned}
\end{equation}

This completes the derivation of our margin loss formulation.

\section{Comparisons with GTE Family}
\label{Appendix C}
\begin{table}[ht]
    \centering
    \scriptsize
    \setlength{\tabcolsep}{2.7pt}
    \caption{Comparative analysis of InfoGain-RAG and various GTE models as rerankers, with Qwen2.5 as the answer generation model on TriviaQA. Results demonstrate InfoGain-RAG's superior performance across all tested configurations.}
    \label{embedding-comparison}
    \begin{tabular}{l|cccc}
    \toprule[1.5pt]
    \textbf{Method} & \textbf{GTE-1.5B} & \textbf{GTE-7B} & \textbf{GTE-Proprietary} & \textbf{InfoGain-RAG} \\
    \midrule[1pt]
    Qwen2.5-0.5B & 45.3\% & 46.5\% & 49.5\% & \textbf{55.8\%} \\
    Qwen2.5-1.5B & 59.7\% & 61.7\% & 63.3\% & \textbf{66.3\%} \\
    Qwen2.5-3B & 63.3\% & 65.6\% & 65.8\% & \textbf{68.2\%} \\
    Qwen2.5-7B & 67.4\% & 69.2\% & 69.5\% & \textbf{72.1\%} \\
    Qwen2.5-14B & 67.5\% & 70.5\% & 71.1\% & \textbf{72.9\%} \\
    Qwen2.5-32B & 70.1\% & 71.9\% & 72.0\% & \textbf{74.7\%} \\
    Qwen2.5-72B & 69.2\% & 72.2\% & 73.4\% & \textbf{76.3\%} \\
    \bottomrule[1.5pt]
    \end{tabular}
\end{table}

\section{Information of Datasets}
\label{Appendix D}

TriviaQA\footnote{\url{https://huggingface.co/datasets/mandarjoshi/trivia_qa}} consists of 174,000 questions based on Wikipedia pages, with answers and their justifications also determined from Wikipedia, including 138,000 for the training set, 17,900 for the validation set, and 17,200 for the test set. NaturalQA\footnote{\url{https://huggingface.co/datasets/sentence-transformers/natural-questions}} is a dataset consists of 307,373 training questions, 7,830 validation questions, and 7,842 test questions. where all questions originate from Google's search records, with answers derived from Wikipedia. PopQA\footnote{\url{https://huggingface.co/datasets/akariasai/PopQA}} contains approximately 14,000 questions all sourced from the Wikidata database. PopQA focuses on long-tail entities and can effectively assess how well a LLM can grasp infrequent factual knowledge. 

FM2\footnote{\url{https://huggingface.co/datasets/tasksource/fool-me-twice/viewer}} is a dataset that contains 10,400 training questions, 1,170 validation qustions and 1,380 test questions, which are designed to test the ability of LLMs to answer simple, factual questions. These questions cover a wide range of topics and are collected from various online sources. The answers to these questions are also provided, making it a valuable resource for training and evaluating question-answering systems.

The December 2018 Wikipedia dump is a comprehensive collection of the content available on Wikipedia up to December 2018. This dump includes nearly 23 millions articles, discussions, and metadata, providing a vast amount of information on a diverse range of topics. It is a valuable resource for natural language processing tasks, such as information extraction, text summarization, and question answering. Researchers and developers can use this dump to train and test their models on a large and diverse corpus of text, helping to improve the performance and accuracy of their systems.

\end{document}